\documentclass[article]{aa}

\usepackage{float}
\usepackage{graphicx}
\usepackage{placeins}
\usepackage{txfonts}
\usepackage{hyperref}
\hypersetup{
    colorlinks,
    citecolor=blue,
    filecolor=blue,
    linkcolor=blue,
    urlcolor=blue
}

\newcommand{\rev}[1]{#1}
\begin{document} 

\title{The disc origin of the Milky Way bulge}

\subtitle{The high velocity dispersion of metal-rich stars at low latitudes}

\titlerunning{Line-of-sight velocity dispersions of stars in the bulge  and their dependence on metallicity}

\author{T. Boin\inst{1}\and
      P. Di Matteo\inst{1}\and
      S. Khoperskov\inst{2}\and
      F. Fragkoudi\inst{3}\and
      S. Ghosh\inst{4,5}\and
      F. Combes\inst{6}\and
      M. Haywood\inst{1}\and
        D. Katz\inst{1}
      }

\institute{GEPI, Observatoire de Paris, PSL Research University, CNRS, Univ Paris Diderot, Sorbonne Paris Cité, Place Jules Janssen, 92195, Meudon, France\\\email{tristan.boin@obspm.fr}
\rev{
\and Leibniz-Institut für Astrophysik Potsdam (AIP), An der Sternwarte 16, 14482 Potsdam, Germany
\and Institute for Computational Cosmology, Department of Physics, Durham University, Durham DH1 3LE, UK
\and Department of Astronomy, Astrophysics and Space Engineering, Indian Institute of Technology Indore, India - 453552
\and Max-Planck-Institut f\"{u}r Astronomie, K\"{o}nigstuhl 17, D-69117 Heidelberg, Germany
\and Observatoire de Paris, LERMA, Collège de France, CNRS, PSL University, Sorbonne University, 75014 Paris, France
}
}

\date{Received XXX; accepted XXX}


\abstract{
Previous studies of the chemo-kinematic properties of stars in the Galactic bulge have revealed a puzzling trend. Along the bulge minor axis,  and close to the Galactic plane, metal-rich stars display a higher line-of-sight velocity dispersion compared to metal-poor stars, while at higher latitudes metal-rich stars have lower velocity dispersions than metal-poor stars, similar to what is found in the Galactic disc. In this work, we re-examine this issue, by studying the dependence of line-of-sight velocity dispersions on metallicity and latitude in APOGEE Data Release~17, confirming the results of previous works. We then analyse an $N$-body simulation of a Milky Way-like galaxy, also taking into account observational biases introduced by the APOGEE selection function. We show that the inversion in the line-of-sight velocity dispersion-latitude relation observed in the Galactic bulge -- where the velocity dispersion of metal-rich stars becomes greater than that of metal-poor stars as latitude decreases -- can be reproduced by our model.  We show that this inversion is a natural consequence of a scenario in which the bulge is  a boxy or peanut-shaped structure, whose metal-rich and metal-poor  stars mainly originate from the thin and thick disc of the Milky Way, respectively. Due to their cold kinematics, metal-rich, thin disc stars are efficiently trapped in the boxy, peanut-shaped bulge, and at low latitudes show a strong barred morphology, which -- given the bar orientation with respect to the Sun-Galactic centre direction -- results in high velocity dispersions that are larger than those attained by the metal-poor populations. Extremely metal-rich stars in the Galactic bulge, which have received renewed attention in the literature, do follow the same trends as those of the metal-rich populations. The  line-of-sight velocity-latitude relation observed in the Galactic bulge for metal-poor and metal-rich stars are thus both an effect of the intrinsic nature of the Galactic bulge (i.e. mostly secular) and of the angle at which we observe it from the Sun.}

\maketitle
%
\section{Introduction}

The Milky Way harbours a boxy, peanut-shaped (hereafter b/p) bulge that has a stellar mass of about $2\times10^{10}M_\odot$ \citep{wegg13}. 
The X-shape signature of the peanut morphology of the bulge can be seen, among other means, through the distribution of red giant stars \citep[e.g.][]{ness12,Li_2012,McWilliam_2010, nataf10, Ness_2016}, whereby the two lobes that constitute the backbone of the peanut shape appear separated by about 1~kpc from the Galactic centre (GC). As several numerical works have shown, a massive classical component -- a bulge that has a spheroidal structure -- can be ruled out beyond a 10\% contribution \citep{Shen_2010, kunder12, Di_Matteo_2014, Di_Matteo_2015, Di_Matteo_2016, gomez18} and the bulge is, for the most part, made of the same stellar populations found in the Galactic (thin and thick) disc, transformed to a b/p morphology via vertical resonance trapping at the bar and buckling instabilities \citep[see the seminal works of][]{combes90, raha91}. Essentially all numerical works developed in the last decade \citep{Shen_2010, bekki11, ness12, ness14, debattista17, athanassoula17, fragkoudi_2017b, fragkoudi18, fragkoudi20}, when compared with data coming from large spectroscopic surveys as ARGOS \citep{freeman13}, BRAVA \citep{howard08}, GIBS \citep{zoccali14}, and APOGEE \citep{majewski17}, and with high-resolution spectroscopic observations \citep{bensby10, bensby11, bensby13, bensby14, bensby17, bensby20, bensby21, nandakumar24}, agree on these conclusions. These observations have indeed provided morphological, kinematic, and chemical abundance relations of bulge stars that constitute stringent constraints with which to compare formation and evolutionary scenarios.

Among these constraints, it has been shown that line-of-sight velocity dispersions (hereafter $\sigma_{v_{los}}$) for populations of stars of different metallicities reveal a surprising trend: along the minor axis and close to the Galactic plane (at  latitudes $|b| \lesssim 5^\circ$), the most metal-rich population displays a higher $\sigma_{v_{los}}$ than the most metal-poor population \citep{Babusiaux_2014, Zoccali_2015, Babusiaux_2016}. This is quite counter-intuitive because, in the Galactic disc, metal-poor stars, which are usually old \citep{fuhrmann98, haywood13, bensby14},  had time to evolve and likely be impacted by several dynamical mechanisms (e.g. interactions with giant molecular clouds in the disc, passages and accretions of satellite galaxies), and hence tend to have a higher velocity dispersion than metal-rich stars, which are usually younger. They can also have been formed kinematically hotter than stars born today. The observed inversion in trends between metal-rich and metal-poor stars observed in these central regions of our Galaxy at latitudes of $|b| \lesssim 5^\circ$ seems to contradict this general trend. This is even more surprising given that at higher latitudes ($|b| \gtrsim 5^\circ$) the trends are those also observed in the disc; that is, the metal-poor populations have the hottest (line-of-sight) kinematics \citep[see, e.g.,][]{Ness_2013}. This raises the question of the possibility of explaining the aforementioned trends in a scenario in which most of the bulge mass is made of disc material, reshaped by the b/p instability. If not, this would suggest a more complex nature of the Galactic bulge than that offered by numerical and theoretical models so far.
    
In the present work, we address this problem by analysing APOGEE Data Release 17 (DR17) observations \citep{Apogee} to quantify the $\sigma_{v_{los}}$-latitude relation along the Galactic bulge minor axis, for stars of different metallicities. We then compare the retrieved relation to the predictions of an $N$-body simulation of a Milky Way-type galaxy, in which the b/p bulge forms out of (thin and thick) disc stars, to test whether such a scenario is able to reproduce the observed trends. As we shall show, such a scenario can satisfactorily explain this apparent paradox. Indeed, because of their cold kinematics, metal-rich, thin disc stars are efficiently trapped in the b/p bulge \citep[as shown by a number of works, see for example,][]{Di_Matteo_2016, Fragkoudi_2017, athanassoula17, debattista17, Ghosh_2024} and, at low latitudes, show a strong barred morphology, which -- given the bar orientation with respect to the Sun-GC direction -- results in high line-of-sight velocity dispersions, larger than those attained by the metal-poor populations. 
 
The paper is organised as follows. In Sect.~\ref{sec:2}, we describe the APOGEE~DR17 data and the selection applied to construct the bulge sample used for the following analysis, as well as the $N$-body simulation and the mock catalogue built from it. In Sect.~\ref{sec:3}, we present the main results of this work; namely, we confirm the existence of an inversion in the  $\sigma_{v_{los}}$-latitude relation along the bulge minor axis, where  metal-rich stars show a steeper gradient than metal-poor stars. This inversion is reproduced well by our $N$-body model and can be explained in terms of a differential response of metal-rich and metal-poor stars to the bar-b/p instability caused by a pre-existing difference in the kinematics of these populations. We show that this inversion strongly depends on the bar viewing angle. Finally, in Sect.~\ref{sec:4} we summarise our results.

\begin{figure*}
          \centering
        \includegraphics[width=0.9\textwidth]{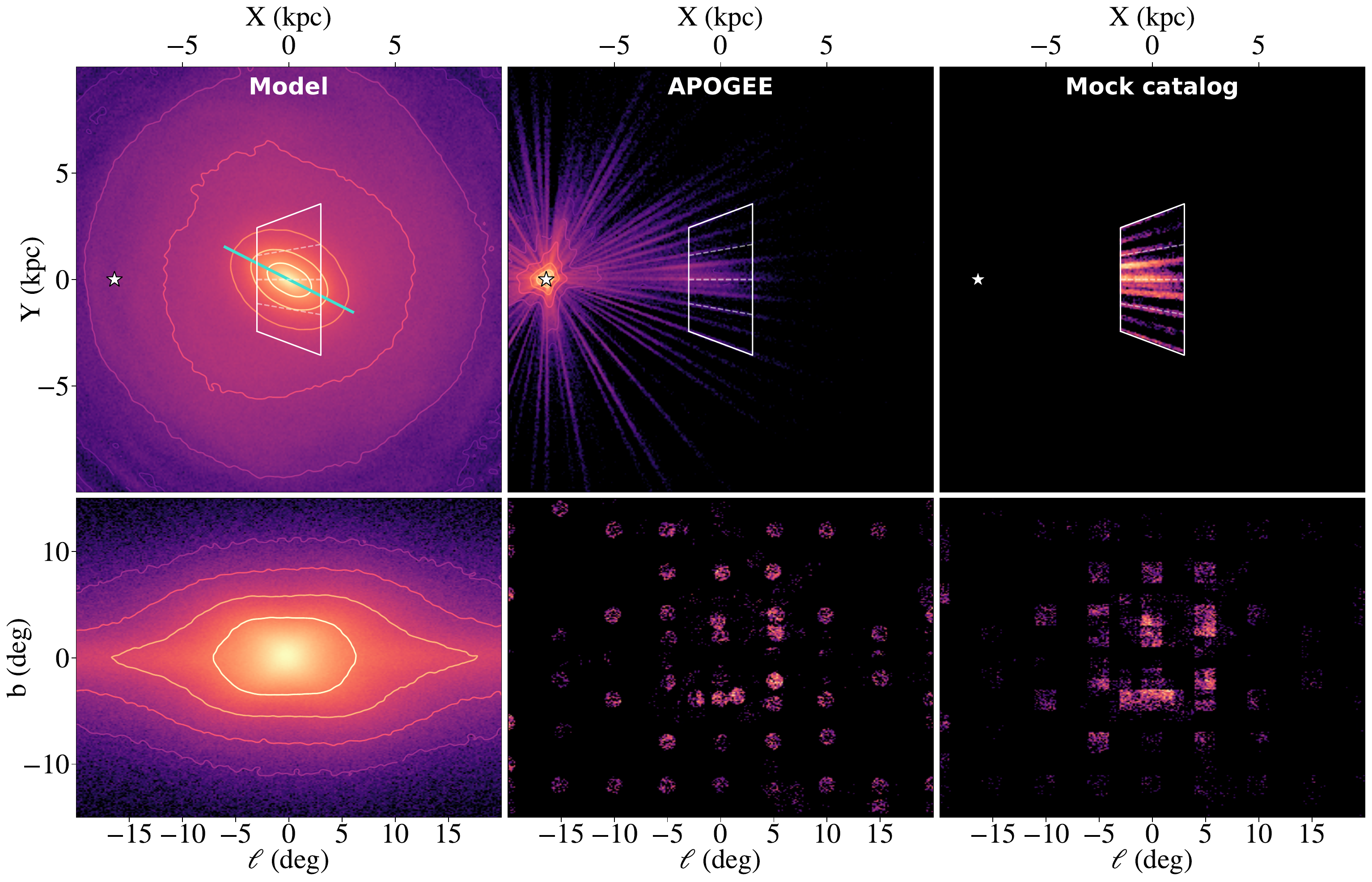} 
        \caption{Density maps of the $N$-body simulation stars \textit{(left column)}, APOGEE stars \textit{(middle column)}, and our mock catalogue stars \textit{(right column)}. \textit{Top panel:} Face-on maps (X-Y-plane), with |X| < 1.5kpc distance bounds and $\pm$ 20$^\circ$ longitude bounds in white, as well as a green line to guide the eye, indicating the angle of the Galactic bar and the Sun at X=-8.12~kpc,Y=0~kpc (white star symbol). \textit{Bottom panels:} Corresponding maps in the Galactic ($\ell$,$b$) plane. The colours indicate relative densities for each of the three catalogues separately.}
        \label{fig:xy_lb}
\end{figure*}

\begin{table*}
    \caption{Initial parameters for each of the components of the $N$-body simulation used in this work.}
    \label{table:1}      
    \centering          
    \begin{tabular}{c | c c c c c c }
        Component & M ($M_\odot$) & r (kpc) & $h_z$ (kpc) & $n_p$ & $\langle[$Fe/H$]\rangle$ (dex) & $\sigma_{[\mathrm{Fe/H}]}$ (dex)\\ 
        \hline
       Thin Disc (\it D1)        & $4.21\times 10^{10}$ & 4.8 & 0.15 & 5000000 &  0.25  & 0.15 \\  
       Intermediate Disc (\it D2) & $2.57\times 10^{10}$ & 2   & 0.3  & 3000000 & -0.26  & 0.2  \\
       Thick Disc  (\it D3)      & $1.86\times 10^{10}$ & 2   & 0.6  & 2000000 & -0.62 & 0.26  \\
       Dark matter halo (r $<$ 100~kpc)  & $1.61\times 10^{11}$ & 10  & -    & 5000000  & -     &   -  \\ 
    \end{tabular}
    \tablefoot{From left to right: total mass, characteristic radius (scale length for the disc components and Plummer radius for the dark matter halo), scale height, number of particles, mean metallicity of the MDF, and dispersion of the MDF. \rev{The model is similar to the M1 model of \citet{fragkoudi_2017b}, with metallicity values taken from \citet{fragkoudi18}.}}
\end{table*}

\section{Observational data and model}\label{sec:2}

\subsection{Selecting bulge stars in the APOGEE catalogue}\label{sec:2-APOGEE}

We have used stars from the APOGEE DR17 catalogue\footnote{\url{https://www.sdss4.org/dr17/irspec}} \citep{Apogee}, which provides spectroscopic quantities, such as line-of-sight velocities and chemical abundances for 647025 stars. This catalogue is complemented by the \textit{astroNN} value-added catalogue \citep{astronn}, which provides distances estimated using a deep-learning approach. From this dataset, we built a bulge sample by filtering stars based on their absolute value of the X-co-ordinate\footnote{In this work, we adopt a galactocentric reference frame, in which the Sun is along the X axis, and disc stars rotate in a clockwise direction.} to the GC ($|X| < 1.5$~kpc), relative heliocentric distance error $\Delta D/D < 0.2$, signal-to-noise ratio  (> 50), and effective temperature, $3 500$ K$ < \rm{T_{eff}} < 5 000$ K. Finally, we cross-matched this catalogue with the APOGEE Value Added Catalogue of Galactic globular cluster stars \citep{schiavon24}, removing stars that have a high probability\footnote{To this aim, we use the membership probability of \citet{vasiliev21} which is also reported in the \citet{schiavon24} catalogue and remove all stars which have a probability higher or equal to 0.9 to be members of one of the Galactic globular clusters.} of being globular cluster members, as we are interested in the analysis of field stars only. From our initial catalogue of 14 632 bulge stars (defined as stars within $|\ell| < 20 ^\circ$, $|b| < 20 ^\circ$ and |X| < 1.5kpc), our final sample contains 5 630 bulge stars.

The resulting coverage of our selected sample is shown in Fig.~\ref{fig:xy_lb} (middle panels), and the star counts in each field in Fig.~\ref{fig:apogee_lb_distri} (right panels). We note that, while some fields have low counts or no stars at all, we have a denser sequence along the Galactic bulge minor axis and close to the Galactic plane at positives latitudes.
    
\subsection{The $N$-body model}

Data from the APOGEE catalogue were compared to a dissipationless $N$-body simulation of a Milky Way-type galaxy, originally introduced in \citet{fragkoudi_2017b}, and further used in \citet{fragkoudi18, fragkoudi19, khperskov19}.\footnote{In this latter work, a higher resolution than that of the ‘standard’ simulation has been employed.} This simulation satisfies a number of properties of the Galaxy, such as  longitudinal and latitudinal metallicity gradients in the bulge, as well as  metallicity distributions and their variations across the bulge extent. While we refer to \citet{fragkoudi18} for a comprehensive description, we recall here its main characteristics. The modelled galaxy is made of four components; namely, a thin, an intermediate, and a thick stellar disc, whose stellar densities follow Miyamoto-Nagai profiles \citep{Miyamoto_1975} characterised  by different masses, scale lengths, and heights, reported in Table \ref{table:1}, as well as a dark matter halo, represented by a Plummer sphere \citep{Plummer_1911}, whose mass and characteristic radius are also reported in Table~\ref{table:1}. The thin, intermediate, and thick discs were modelled with 5, 3, and 2 million particles, respectively, and the dark matter halo is represented by 5 million particles (see Table~\ref{table:1}).

The initial positions and velocities were set by employing the iterative method described in \citet{rodionov_2009}, by building an equilibrium phase model such that the resulting density distribution matches the constraints given by the density profiles we imposed on each component. The simulation was run using a parallel MPI Tree-code~\citep{Khoperskov_2014}, with the gravitational forces being computed with a tolerance parameter of $\theta=0.7$ and an $\epsilon = 50$ pc softening was used. The equations of motion were integrated with a timestep, $\Delta t = 0.2~$Myr.

The thin disc is associated with a metal-rich ([Fe/H] > 0) and kinematically cold initial population of stars (referred to as \textit{D1} in the original paper), the intermediate disc is associated with an intermediate metallicity (-0.5 < [Fe/H] < 0) and intermediate initial velocity dispersion population (\textit{D2}), and the thick disc is associated with a metal-poor ([Fe/H] < -0.5) and kinematically hot initial population (\textit{D3}), and thus mimics the main properties of the Galactic disc \citep[see][]{haywood13}. In particular, the scale heights and scale lengths of these three discs are compatible with the findings by \citet{bovy12}, even if here we discretise, with three components only, a disc where populations of different metallicities have a more continuous distribution. Disc particles were assigned a metallicity by drawing randomly from a normal distribution, where each disc has a constant (i.e. without any radial or vertical metallicity gradient within each component) mean metallicity and dispersion (see Table \ref{table:1}).

The simulation was evolved for 7 Gyr, well after the bar (t=0.5~Gyr) and b/p bulge (t=5~Gyr) formed, and in this work we have analysed one of its last snapshots, at t=7~Gyr. In order to make a proper comparison with the observational data, we re-scaled the positions of this final snapshot by a factor of 0.5, so that the semi-major axis of the simulated bar has a length of 5~kpc, which is compatible with observations \citep{Wegg_2015}. We consequently also re-scaled the masses, and time, to preserve the virial equilibrium of the system. To build the Galactic frame as well as to compute line-of-sight velocities, the Sun was placed along the X axis, at a distance of 8.12~kpc from the centre of the simulated galaxy, and at an angle of $27^\circ$ with respect to the bar, so as to match the position of the Sun relative to the centre of the Milky Way \citep{Babusiaux_2005,wegg13,Wegg_2015, Chatzopoulos_2015, Reid_2014, Cao_2013, Rattenbury_2007, Stanek_1997,Lopez_2005}. The density maps of the analysed snapshot are reported in  Fig.~\ref{fig:xy_lb} (left panels), where projections both on the galaxy plane and in galactic co-ordinates are shown. 

 \begin{figure*}
    \centering
    \includegraphics[width=0.5125\textwidth]{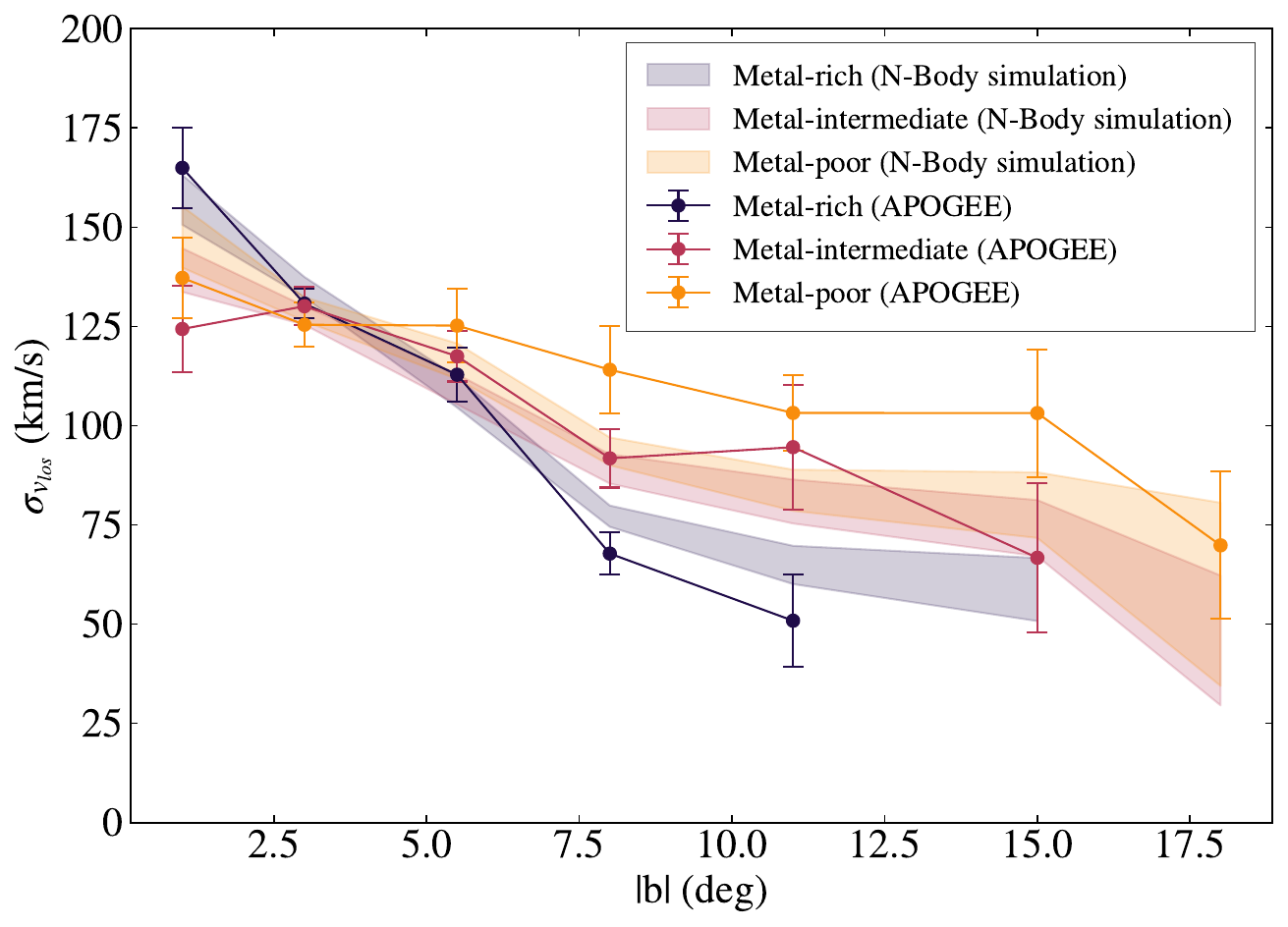}
      \includegraphics[width=0.48\textwidth]{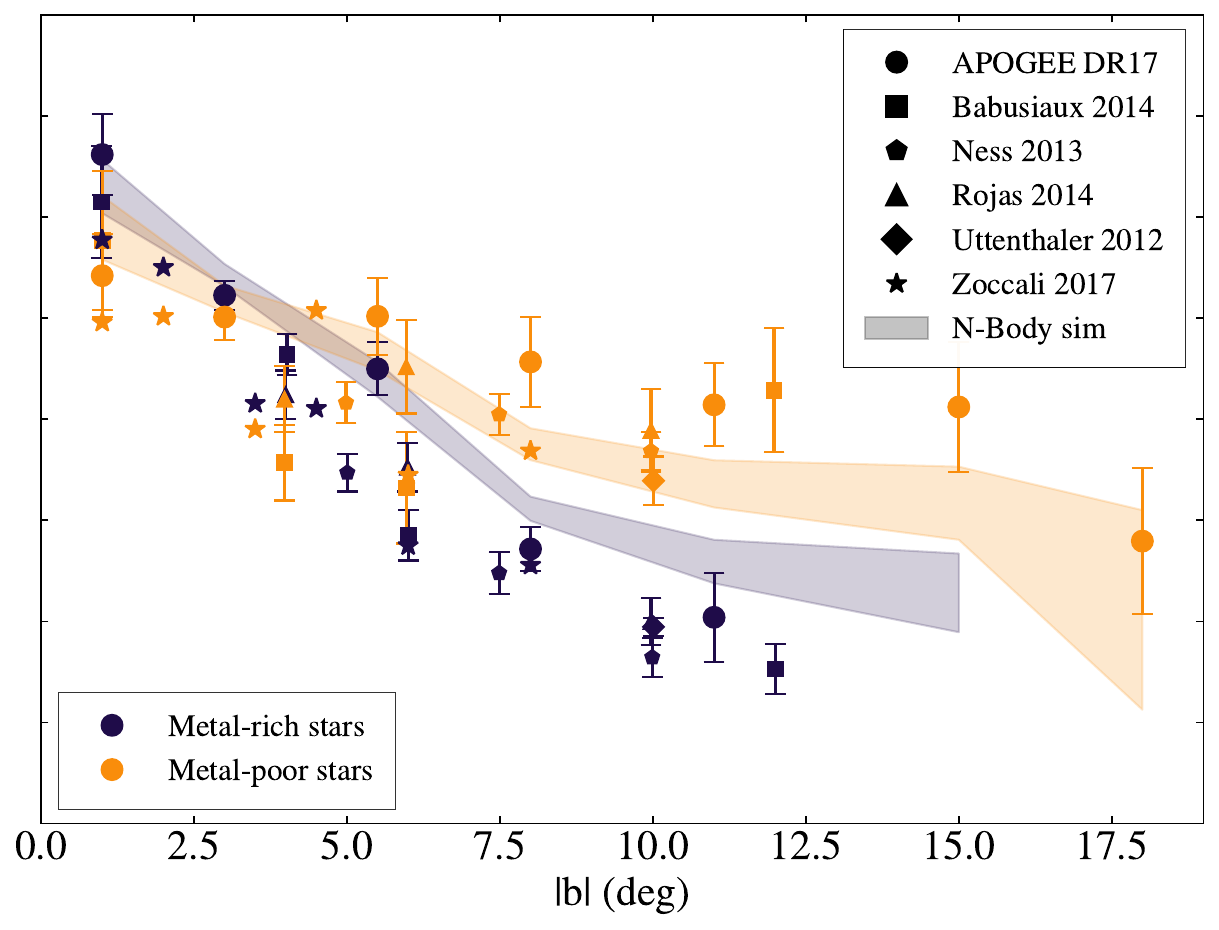}
          \caption{Line-of-sight velocity dispersion, $\sigma_{v_{los}}$, profiles. \textit{Left panel}: our quality APOGEE sample (\textit{error bar curves}) and our mock catalogue (\textit{shaded areas of height $\pm \sigma$}). The profiles are plotted for stars with |$\ell$|$<2^\circ$ and |X| $<$ 1.5kpc. \textit{Right panel}: profiles of our quality APOGEE sample (\textit{error bar with circle symbol}) and of our mock catalogue (\textit{shaded areas of height $\pm \sigma$}), compared to results from \cite{Babusiaux_2014} and \cite{Babusiaux_2010} (\textit{squares}), \cite{Ness_2013} (\textit{pentagons}), \cite{Rojas_2014} (\textit{triangles}), \cite{Uttenthaler_2012} (\textit{diamonds}), and \cite{Zoccali_2015} (\textit{stars}).}
    \label{fig:disp_apogee_sim}
\end{figure*}

Finally, similarly to \citet{fragkoudi18}, to make a robust comparison of the model to the observational data, we take into account the APOGEE selection function by imposing the APOGEE (heliocentric) distance distribution function (DDF), along different lines of sight, on the model.  The details of the procedure adopted to reproduce the DDF  for each line of sight are described in Appendix~\ref{mock}, and the resulting mock catalogue, where the simulation has been convolved with the APOGEE DDF, is shown in Fig.~\ref{fig:xy_lb}, for the projection on the Galactic plane (X-Y) and in ($\ell - b$) co-ordinates (top and bottom panels, respectively).

\section{Results}\label{sec:3}

\subsection{Comparing APOGEE~DR17 and literature data with the $N$-body model}

 In Fig.~\ref{fig:disp_apogee_sim} (left panel), the velocity dispersion profiles of metal-rich ($\rm{[Fe/H]} \ge 0$), metal-intermediate ($-0.5 \le \rm{[Fe/H]} < 0$), and metal-poor ($\rm{[Fe/H]} < -0.5$) populations along the bulge minor axis are shown, for the $N$-body model and the APOGEE data.
 In both cases, we have selected only  stars with longitudes of |$\ell$| < 2°   and with a galactocentric co-ordinate, |X| < 1.5~kpc.
  To construct the velocity dispersion profiles from the simulation, we have applied the APOGEE DDF to the model, as was mentioned in the previous section, and as is described more in detail in Appendix~\ref{mock}. As for the APOGEE data, we selected bulge stars using the criteria described in Sect.~\ref{sec:2-APOGEE}. Both in the observational and simulated data, we used a wide latitude binning to mitigate low star counts by grouping neighbouring fields (where we used the following latitude ranges for the bins : ([0°,2°], [2°,4°], [4°,7°], [7°,9°], [9°,13°], [13°,17°], and [17°,19°]), and uncertainties in the velocity dispersion profiles were computed via a bootstrapping method. 
 
Comparing the velocity dispersion profiles along the bulge minor axis of metal-rich, metal-intermediate, and metal-poor stars obtained by analysing the APOGEE bulge sample (solid lines in Fig.~\ref{fig:disp_apogee_sim}, left panel), it is evident that for   $|b | \gtrsim 5^\circ$ the trend with metallicity is such that the lower the metallicity, the higher the velocity dispersion, as is expected if stars in these metallicity ranges arise from populations whose kinematics becomes hotter with decreasing metallicity \rev{(and as is shown by the profiles at t=0~Gyr in Appendix~\ref{ic}, where the discs are still in their initial conditions and no bar or bulge has formed)}. At $|b| < 5^\circ$ the trend is reversed, and the metal-poor populations become the kinematically coolest.
The $N$-body simulation shows the same trend: while at high absolute latitudes (  $|b | > 5^\circ$ ) the metal-rich stars are, among the three populations analysed, those characterised by the lowest line-of-sight velocity dispersions, as $|b |$ diminishes, they become the hottest population, by reaching $\sigma_{vlos} \sim 160$~km s$^{-1}$ close to the Galactic mid-plane.

In Fig.~\ref{fig:disp_apogee_sim} (right panel), we compare the dispersion profiles resulting from the  APOGEE bulge sample, and from our mock catalogue, with the velocity dispersion profiles along the bulge minor axis obtained by previous works. In this figure, stellar particles of the $N$-body simulation have been divided into two groups only, metal-rich ($\rm{[Fe/H]} \ge 0$) and metal-poor ($\rm{[Fe/H]} < -0.5$), and are compared with data from the literature, where often bulge stars are grouped into two components only \citep[but see][for a finer subdivision of the metallicity range]{Ness_2013}. Literature data points have been extracted from \citet{Babusiaux_2016} (and are originally from \citealt{Babusiaux_2010}, \citealt{Babusiaux_2014}, \citealt{Ness_2013}, \citealt{Rojas_2014}, and \citealt{Uttenthaler_2012}), to which we added data from \citealt{Zoccali_2015}. 

In general, we see that the results obtained by analysing the APOGEE data are, within the uncertainties, in agreement with previous works: close to the Galaxy midplane, metal-rich bulge stars have a higher line-of-sight velocity dispersion than metal-poor stars, and a clear inversion to this trend is seen for $|b| \ge 5^\circ$. As a result of this inversion, the slope of the $\sigma_{vlos}-|b|$ relation for metal-rich stars is steeper than that of the metal-poor stars: 
for $|b| \le 7.5^\circ$, where most of the observational data are found, the two slopes are, respectively, $-15.99\pm 1.39$~km s$^{-1}$ deg$^{-1}$ and $-8.07\pm 2.27$~km s$^{-1}$ deg$^{-1}$. In the same latitude range, the slopes found in the simulations are similar, with $-10.62\pm 0.15$~km s$^{-1}$ deg$^{-1}$ for metal-rich stars and $-7.33\pm 0.05$~km s$^{-1}$ deg$^{-1}$ for metal-poor stars. We mention that, in calculating these slopes, we have given a similar weight to all observational data points, despite the fact that they do not all use the same definition for metal-rich and metal-poor stars (while the compilation of \citet{Babusiaux_2016} adopts a separation at $ 0 < \rm{[Fe/H]} < 0.5$ and at  $-1 < \rm{[Fe/H]} < -0.5$, respectively,   \citet{Zoccali_2015} adopts a slightly different definition for metal-rich stars ($\rm{[Fe/H]} > 0.1$ in their work) and a broader interval for metal-poor stars ($\rm{[Fe/H]} < -0.1$)\footnote{Note also that in  \citet{Zoccali_2015}, uncertainties in the calculated velocity dispersions are not reported, hence the lack of error bars in the associated points in  Fig.~\ref{fig:disp_apogee_sim} (right panel).}). Despite the differences in the definitions of the two groups, as well as the use of different distance estimates for stars in the analysed samples, the different trends in the kinematics of the two groups are visible. Indeed, the fact that they are present irrespective of the details of the analysis (i.e. metallicity range, distance estimates, amplitude of the longitude and latitude intervals) strengthens the robustness of these findings.

\begin{figure}
    \centering
    \includegraphics[width=0.5\textwidth]{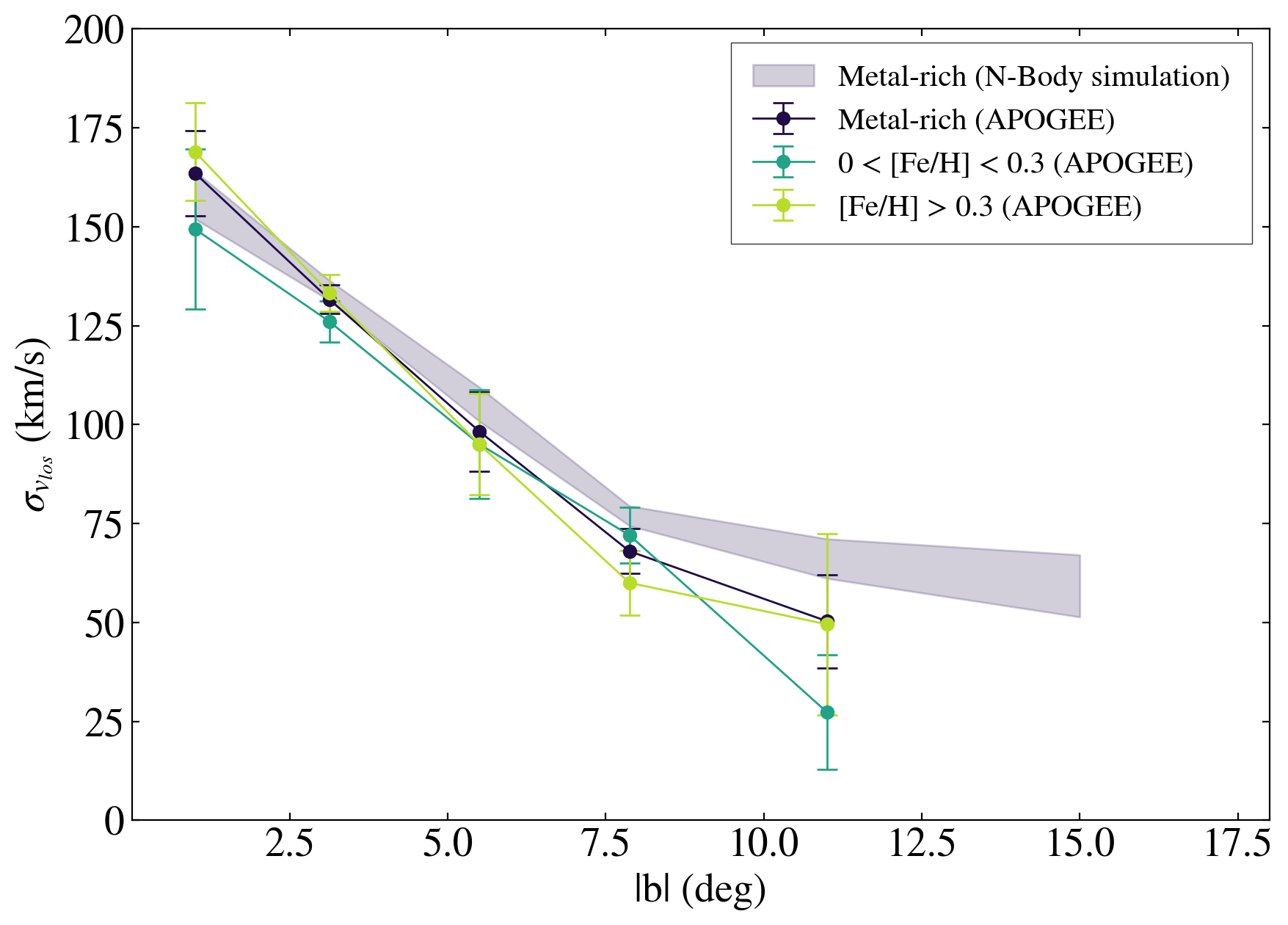}
          \caption{$\sigma_{v_{los}}$ line-of-sight velocity dispersion profiles of our quality APOGEE sample (\textit{error bar curves}) and of our mock catalogue (\textit{shaded areas of height $\pm \sigma$}). The profiles are plotted for stars with |X| $<$ 1.5kpc and |$\ell$|$<2^\circ$, for metal-rich stars ([Fe/H] > 0), and two metal-rich populations defined as very metal-rich (0 < [Fe/H] $\leq$ 0.3) and extremely metal-rich ([Fe/H] > 0.3).}
    \label{fig:EVMR}
\end{figure}

Finally, it is worth citing in this context some new results about the presence of extremely metal-rich stars in the Galactic bulge, whose existence has been known for some time \citep[see, for example, ][]{fulbright06, johnson07, Zoccali_2008}. \citet{rix24} recently suggested  that this population forms a  dynamically hot system, a ‘knot’, in the inner 1.5~kpc of our Galaxy \citep[see also][]{horta24}. While we defer a specific study of these stars to future works, it is nevertheless interesting, in the context of the present paper, to investigate the $\sigma_{vlos}$-|$b$| relation that they trace, because the latter helps to constrain their nature.
To this aim, in Fig.~\ref{fig:EVMR}, we have split the APOGEE metal-rich bulge population into two groups, made, respectively, of 277 stars with $0< \textrm{[Fe/H]} \leq 0.3$ (hereafter called very metal-rich (VMR) stars), and  of 272 stars with $0.3 < \textrm{ [Fe/H]}$ (hereafter called extremely metal-rich (EMR) stars). The figure shows that these groups share the same $\sigma_{vlos}$-|$b$| relation, which points to the fact that:
1) at low latitudes, EMR stars are as dynamically hot (in line-of-sight projection) as VMR stars, and as the metal-rich population as a whole; 2) similarly to VMR stars, the EMR population shows a steep $\sigma_{vlos}$-|$b$| gradient, with $\sigma_{vlos}$= 55 km~s$^{-1}$ at $|b| \sim 10$ $^\circ$, and rising up to $\sigma_{vlos}$= 170 km~s$^{-1}$ at $|b| < 2$ $^\circ$. These findings suggest that, in terms of line-of-sight kinematics, EMR stars are not exceptional, but share the same kinematic behaviour as the bulk of the metal-rich population ($0< \rm{[Fe/H]}< 0.3$). We note that the choice of separating VMR and EMR at a 0.3~dex metallicity was done to maximise the number of EMR stars, which rapidly decreases with increasing metallicity, and thus our choice of definition of VMR and EMR stars does not exactly match the one made in \citet{rix24}.

\subsection{Understanding the inversion in the line-of-sight velocity dispersions at low latitudes}

\begin{figure*}
    \centering
    \includegraphics[width=.46\textwidth]{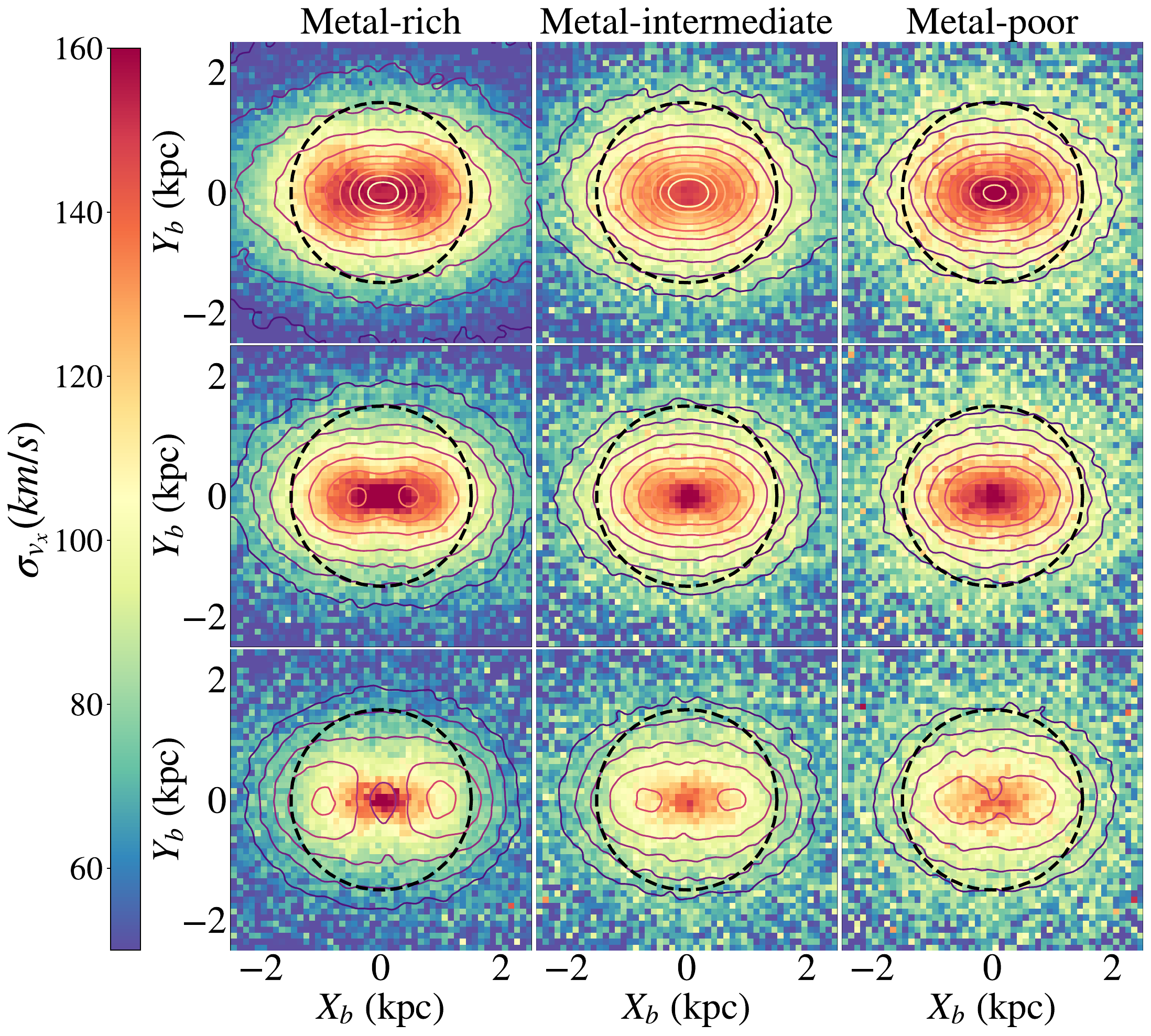}
    \includegraphics[width=.4515\textwidth]{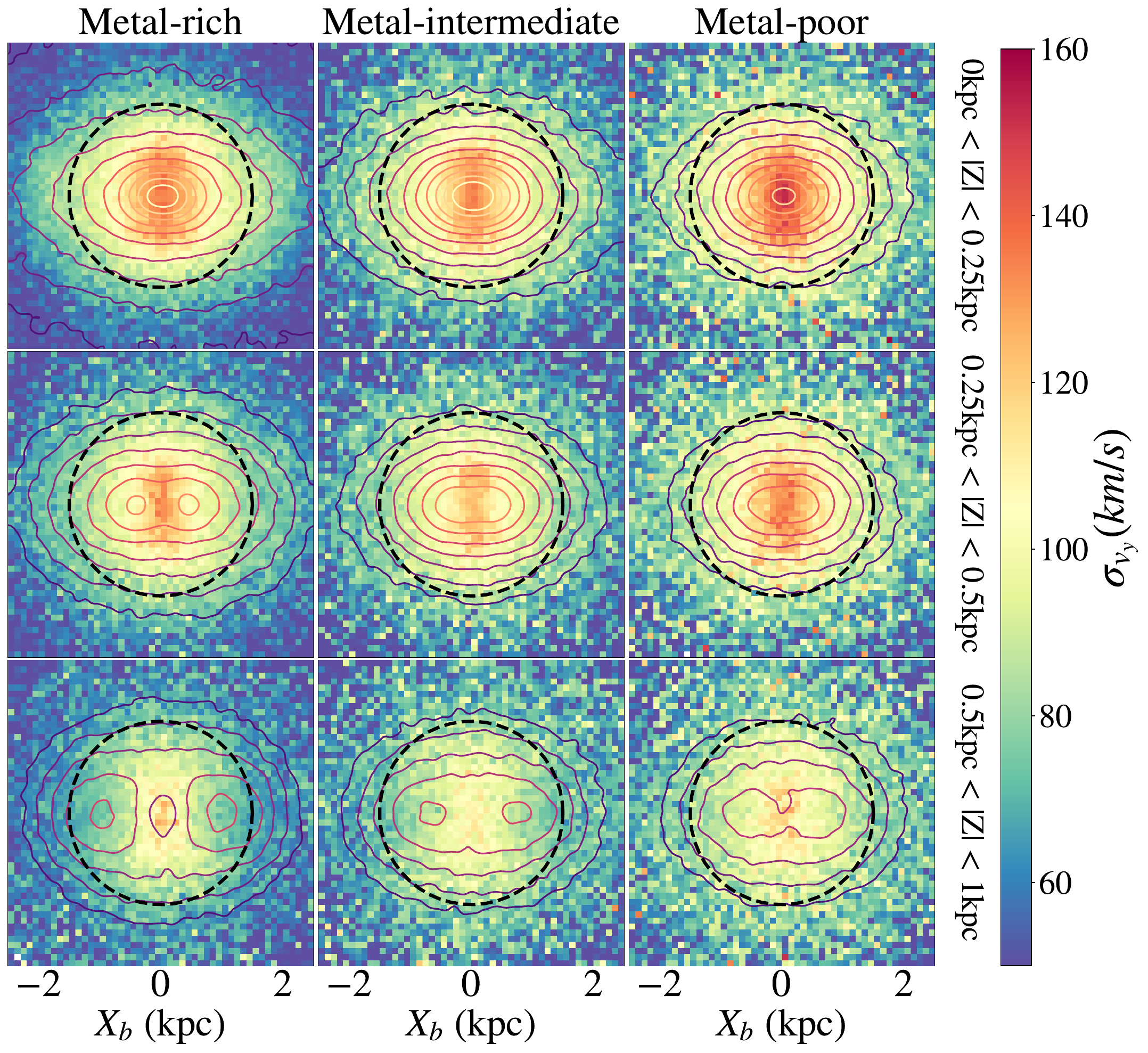}
    \caption{Maps of the velocity dispersion maps for the bar-oriented $\sigma_{v_{Xb}}$ (\textit{left panel}) and $\sigma_{v_{Yb}}$ (\textit{right panel}) components. Each population (metal-rich, metal-intermediate, and metal-poor) correspond to a column in each panel, and different height cuts are represented on each row (0~kpc < |Z| < 0.25~kpc, 0.25~kpc < |Z| < 0.5~kpc, and 0.5~kpc < |Z| < 1~kpc). The axes are bar-oriented, isodensity contours are shown as coloured lines, and a 1.5~kpc region around the GC is shown as a dotted black circle.}
     \label{fig:xy_v}
\end{figure*}

 The comparison of observational data with the $N$-body simulation presented in the previous section shows that even a model in which bulge populations with $\rm{[Fe/H]} \ge -1$ originate exclusively from disc stars is able to reproduce the observations, and in particular the apparent ‘dual’ nature of the metal-rich population, which is kinematically hot at low latitudes and kinematically cold at higher latitudes. As a further step, we need to understand why, in the model, bulge metal-rich stars -- originally belonging to the kinematically coldest disc (i.e. \textit{D1}) -- end up showing the largest line-of-sight velocity dispersions at low latitudes.
 
To this aim, in Fig.~\ref{fig:xy_v} we show the maps of the velocity dispersion of metal-rich, metal-intermediate, and metal-poor stellar particles of the $N$-body model, projected on the galaxy midplane, for two different values of the dispersion measured, respectively, along the in-plane bulge major ($\sigma_{v_{Xb}}$, left panels) and in-plane minor ($\sigma_{v_{Yb}}$, right panels) axis. In this figure, we have chosen a galactocentric reference frame in which the $X_b$ axis is parallel to the bar major axis, and the $Y_b$ axis is perpendicular to it, with both axes lying on the galaxy midplane. Different rows in  Fig.~\ref{fig:xy_v} show different slices in height from the midplane for stars in the three metallicity intervals at, respectively, $|Z| \le 0.25$~kpc (|b| $\le \sim 1.8$° at the GC), 0.25~kpc$ < |Z| \le 0.5$~kpc ($\sim 1.8$°$\le$ |b| $\le \sim 3.6$° at the GC), and  0.5~kpc$ < |Z| \le 1$~kpc ($\sim 3.6$°$\le$ |b| $\le \sim 7.1$° at the GC).
Density contours are also reported in all the plots to link kinematic properties to morphological features. Similar maps, but in the ($\ell,b$) plane and for a Sun-Galactic Bar angle of 27°, are shown in Appendix~\ref{lb_v}.

As for the morphological properties, from this figure we can see that:
\begin{itemize}
\item The b/p structure is visible at larger heights above the plane ($0.5 < |Z| \le 1$~kpc) as a double-lobe feature, or split, in the density contours. At these heights above the plane, this feature is stronger for the metal-rich population and weaker for the metal-intermediate one, and absent for the metal-poor one. The b/p structure disappears at smaller |Z|. These trends are all qualitatively similar to those found in other $N$-body simulations \citep[among others, see][]{Fragkoudi_2017,fragkoudi18}, as well as in observational data, where the presence of the  b/p structure is traced, for instance, by the split in the distribution of red clump stars \citep[see, e.g.][]{Ness_2013a,Ness_2013}.
\item Apart from the double-lobe feature in some of the isodensity contours, most of them appear nearly elliptical, the ellipticity being higher for the metal-rich population than for the metal-poor one (see Appendix~\ref{eccentricity} for more details on this point). This difference in ellipticity of the isodensity contours reflects a difference in the strength of the bar, as traced by metal-rich, metal-intermediate, and metal-poor populations, in agreement with a number of previous studies \citep[see][]{Babusiaux_2010,Fragkoudi_2017,debattista17,Bekki_2011,Athanassoula_2009, combes90}.
\end{itemize}

As for the kinematic properties, we can see that:
\begin{itemize}
\item The structure and amplitude of the maps of the velocity dispersion along the bar major axis ($\sigma_{v_{Xb}}$)  vary both as a function of metallicity and height above the plane.  
\item Close to the galaxy midplane ($|Z| \le 0.25$~kpc), metal-rich stars show velocity dispersions with a butterfly-like pattern, characterised by high values (greater than 140~km s$^{-1}$) over most of the inner 1.5~kpc region of the simulated galaxy. At $|Z| \le 0.25$~kpc, the intermediate population shows the lowest values of $\sigma_{v_{Xb}}$, while the metal-poor population has $\sigma_{v_{Xb}}$ values similar to that of the metal-rich one, but extended over a more limited region than the metal-rich particles (compare the spatial extension of the red-brown region in the left and right columns of Fig.~\ref{fig:xy_v}, left panels).
\item At higher distances from the galactic plane, in the innermost few hundred parsecs from the galaxy centre, the metal-rich population sustains a higher  $\sigma_{v_{Xb}}$ than the metal-intermediate and metal-poor populations whose central velocity dispersion decreases with |Z|. However,  except for this innermost region, one sees that, overall, the $\sigma_{v_{Xb}}$ values are larger for the metal-intermediate and metal-poor stars than for the metal-rich ones (while the former have $\sigma_{v_{Xb}}$ of about 100~km s$^{-1}$ inside most of the  region delimited by R=1.5~kpc,  the $\sigma_{v_{Xb}}$ of the metal-rich population rapidly decreases to values as low as $\sim 80$~km s$^{-1}$ in the same spatial region).
\item The velocity dispersions perpendicular to the bar major axis ($\sigma_{v_{Yb}}$, see Fig.~\ref{fig:xy_v}, right panels) show a different behaviour, with the metal-rich population always being colder than the metal-intermediate and metal-poor ones, at low vertical distances as well as at large vertical distances to the galaxy midplane.
\end{itemize}

\begin{figure*}
    \centering
    \includegraphics[width=\textwidth]{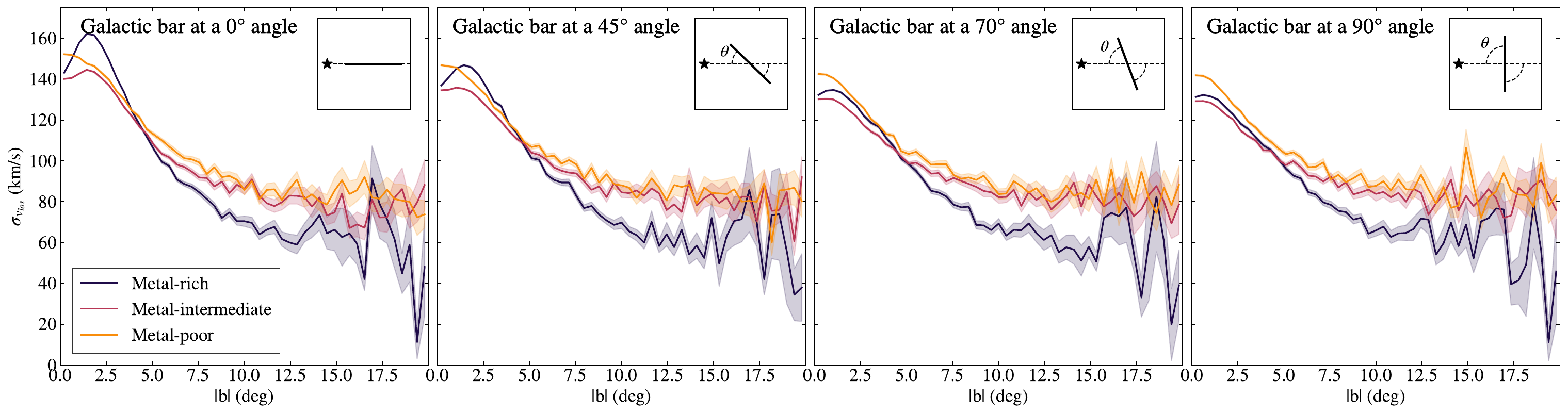}
      \caption{Line-of-sight velocity, $\sigma_{v_{los}}$, dispersion profiles of simulation particles for a Galactic bar angle relative to the Sun of, from left to right, 0°, 45°, 70°, and 90°. The profiles are plotted for stars with |$\ell$|$<2$°  and |X| $<$ 1.5~kpc, and the shaded areas show a $\pm \sigma$ uncertainty. Inset plots show the position of the Sun relative to the bar orientation.}
    \label{fig:dispersion_0_90}
\end{figure*}

From these findings, we can understand that the asymmetry in the mass distribution of the inner disc regions (bar and b/p bulge) is also accompanied, in these same regions, by an asymmetry in the kinematics of the stars. The velocity dispersion generally has a higher intensity along the major axis of the bar than along the direction perpendicular to it (i.e. implying velocity dispersion maps that are also asymmetrical). Its intensity also varies with the metallicity of the stellar population, and this trend is a natural consequence of a bar formed from a disc in which a pre-existing relation between the metallicity of the stars and their velocity dispersion, similar to that observed in the Galactic disc \citep[e.g.][]{bovy12, haywood13}, was in place. In this case, as has been discussed in previous works, kinematically colder and thinner stellar discs produce stronger bars with greater ellipticity than kinematically hotter and thicker discs \citep{Bekki_2011, Fragkoudi_2017,athanassoula17, debattista17, Ghosh_2023}.
    
The results presented in Fig.~\ref{fig:xy_v} have a natural consequence: the  line-of-sight velocity dispersion of stars in a bar-b/p structure changes with the viewing angle.  To further address this point, in Fig.~\ref{fig:dispersion_0_90} we show the $\sigma_{v_{los}}-|b|$ relation for metal-rich ($\textrm{[Fe/H]} > 0$), metal-intermediate (-0.5 < [Fe/H] < 0), and metal-poor ([Fe/H] < -0.5) particles, in the cases of a simulated bar observed end-on (i.e. the line-of-sight being parallel to the bar major axis and the bar viewing angle being equal to $0^\circ$), \rev{side-on}  (i.e. the line-of-sight being perpendicular to the bar major axis and the bar viewing angle being equal to $90^\circ$) and at intermediate angles of $45^\circ$ and $70^\circ$. The variation of $\sigma_{v_{los}}$ with latitude significantly changes in the four cases, as is anticipated from Fig.~\ref{fig:xy_v}. If the bar is viewed \rev{side-on}, the resulting line-of-sight velocity dispersion has a monotonic trend with metallicity, at all $|b|$, with the bulge metal-rich population having lower values $\sigma_{v_{los}}$ than the bulge metal-poor population, at all latitudes. If the orientation of the bar angle is intermediate ($45^\circ$), an inversion in the $\sigma_{v_{los}}$ values for metal-rich and metal-poor components starts to appear at $|b|\sim 5^\circ$. When the bar is viewed end-on, the  $\sigma_{v_{los}}$  of the metal-rich population at $|b|\sim 2- 5^\circ$ is definitely higher than that of more metal-poor stars. Interestingly, at $|b| \lesssim 2^\circ$, a drop in the line-of-sight velocity dispersion of metal-rich stars is predicted by our $N$-body model. The origin of this feature needs further investigation, which requires higher-resolution simulations than the one presented in this paper.
We note that in Fig.~\ref{fig:dispersion_0_90} we used all particles of the $N$-body simulation without any sampling or convolution with DDF, since in this figure we are not comparing the model with any observational data. Profiles made with continuously increasing bar--observer (i.e. Sun) position angles between 0 and 90$^\circ$ show a continuous trend of metal-rich stars becoming hotter close to the plane (with the specific configuration of an observer placed at an angle of $\sim 27^\circ$ relative to the Galactic bar being represented in Fig.~\ref{fig:disp_apogee_sim}). This certainly does not suggest that the kinematics of the populations inherently change with the position of the observer, but rather that the velocity dispersions we are referring to are measured through the line-of-sight component.

We interpret this dependence of the dispersion inversion on the Sun-Galactic bar orientation as follows: at low angles, the line-of-sight velocities in the innermost regions of the bulge are mainly dominated by the $v_{Xb}$ component of velocities, the component along the major axis of the bar. As the dispersion maps in Fig.~\ref{fig:xy_v} show, the metal-rich population displays a wider region of high $\sigma_{v_{Xb}}$ dispersion than the metal-poor population. Hence, at low Sun-Galactic bar angles, we expect the line-of-sight velocity dispersion profiles to reflect the difference in $\sigma_{v_{Xb}}$ dispersions. At angles closer to 90°, however, the line-of-sight velocity is mainly dominated by the $v_{Yb}$ component (perpendicular to the bar major axis), for which the dispersion maps show higher dispersion for metal-poor stars than for metal-rich ones. Thus, the right panel of Fig.~\ref{fig:dispersion_0_90} shows no or a very weak dispersion inversion between the metal-rich and metal-intermediate stars. This can also be seen in the dispersion maps of Fig.~\ref{fig:xy_v}, for the $\sigma_{v_{Yb}}$ component. Indeed, we observe, at low |Z| (top panels), that the metal-rich and metal-intermediate populations display similar patterns of dispersion in their central regions. At higher latitudes ($|b| > 5^\circ$), we find that the trends do not depend on the Sun-bar angle anymore but rather display the expected result of the metal-poor population being kinematically hotter than the metal-rich one. In fact, dispersion maps at higher Z-slices show that the dispersion patterns seen in Fig.~\ref{fig:xy_v} near the Galactic plane are not present at high altitudes, highlighting the fact that the dispersion inversion is a near-Galactic plane effect of the kinematics of each components. Paradoxically, it is the fact that the metal-rich population is kinematically colder than the metal-poor population that causes it to be more efficiently trapped in Galactic plane-bound and eccentric orbits, and thus leads to a higher line-of-sight velocity dispersion being generated at low latitudes.    

\subsection{Limitations of the $N$-body model and possible effects on the $\sigma_{vlos}$-|$b$| relation}

The $N$-body model with which observational data are compared lacks at least two ingredients that may have an impact on the $\sigma_{vlos}$-|b| relation.
\begin{enumerate}
\item A gaseous disc, and associated star formation. In our $N$-body model, the whole thin disc is present since the beginning of the simulation, and thus does not form stars over time. As a consequence,  metal-rich, thin disc stars all have the same velocity dispersion, before the bar and b/p bulge form. The presence of ongoing star formation and the birth of young stellar populations over time should have the consequence of reducing the velocity dispersion of the disc, possibly counteracting its kinematic heating over time. In the comparison shown in Fig.~\ref{fig:disp_apogee_sim}  (left and right panels), metal-rich particles in the $N$-body simulation tend to have a slightly higher $\sigma_{vlos}$ than in the APOGEE data, at $|b| \gtrsim 5^\circ$. We shall investigate in future works whether the reason of this difference -- which is within 1-2~$\sigma$ from the data -- may be due to a lack of continuous star formation in the model.
\item  A stellar halo. A number of studies have shown that the stellar halo of the Milky Way is composite \citep[e.g.][]{nissen10, haywood18, dimatteo19, belokurov20, naidu20, horta23}, with an accreted and an in situ population, both of which are kinematically hotter than the Galactic thin and thick discs . Even if the stellar halo is a tiny fraction of the total stellar mass budget in the Galaxy, halo stars with [Fe/H]$ > -1$ in the inner regions of the Milky Way may contribute to increasing the overall velocity dispersions of the most metal-poor  components, and thus, in particular, their $\sigma_{vlos}$. The comparison made in Fig.~\ref{fig:disp_apogee_sim} (left panel) between the APOGEE data and the $N$-body simulation shows that the former tend to show a higher $\sigma_{vlos}$ than the latter, at all latitudes $|b| > 5^\circ$. It is, however, premature to ascribe this difference to the absence of a halo component in the simulation, which may, in turn, affect the observational estimates. Indeed, Figure~\ref{fig:disp_apogee_sim} (right panel) shows that the difference between the $N$-body model and the APOGEE data is of the same order as the difference found between APOGEE data and other observational studies \citep[see, for example,][]{Ness_2013, Zoccali_2015}.
\item Finally, the reader may have noted that no classical bulge is included in this simulation. As was recalled in the introduction, several works have shown that -- if present -- a classical bulge should represent a few percent of the stellar mass of the Milky Way (so, typically, a few $10^8 M_\odot$) and no more than one tenth (so $\sim 5 \times 10^9 M_\odot$). \rev{\citet{Shen_2020} discussed the incompatibility of the presence of a massive classical bulge in the centre of the Milky Way with the observations of a distinct nuclear star cluster, which would have been impossible to detect were this classical bulge significantly massive}. Such a classical component, according to the mass-size relations of spheroids \citep[e.g. see][their Fig.~4]{hon23}, should have an effective radius of less than 200 pc at most, which translates into an angular size of 1.4 $^\circ$ at the distance of the GC. So, even if it were present, this classical component would have no effect on the inversion observed at $b \sim 5$ $^\circ$ in the $\sigma_{vlos}$-|b| relation. Its presence may affect the velocity dispersion of the most metal-poor stars in our analysis \citep[because of the existence of a mass-metallicity relation for spheroids, classical bulges of $10^8-10^9 M_\odot$ would have a typical mean metallicity between $-1$ and $-0.5$~dex, see][]{kirby13, dominguez23} only in the innermost 1-2 degrees of the Galaxy. In this respect, it may be interesting to investigate to what extent such a small, metal-poor component can affect the $\sigma_{vlos}$ at low latitudes, and -- by making use of the comparison with observational data -- finally prove (or disprove) its presence in the inner Milky Way.
\end{enumerate}
    
\section{Conclusions}\label{sec:4}

In this work, we have made use of APOGEE~DR17 data  to derive the line-of-sight velocity dispersion of  stars along the bulge minor axis (i.e. as a function of latitude $b)$ for different metallicity bins. Our results confirm previous works \citep{Babusiaux_2014,Babusiaux_2016,Rojas_2014, Zoccali_2015} about the existence of an inversion in the  amplitude of the line-of-sight velocity dispersion of metal-rich and metal-poor stars, with the former having a higher dispersion at $|b| < 5^\circ$ than metal-poor stars.
We analysed a previously developed $N$-body simulation where the bar-b/p structure forms out of a disc, where metal-rich stars are distributed in a thin disc, and metal-poor stars in a thicker disc component. This allowed us to demonstrate that the trends observed in the observational data are a natural consequence of the differential mapping of kinematically cold and hot disc populations into a bar and b/p bulge.
Metal-rich, thin disc stars are trapped efficiently in the bar instability, which results in  high velocity dispersions along the bar major axis, especially close to the Galaxy midplane. A steep velocity dispersion gradient ($-15.99\pm 1.39$~km s$^{-1}$ deg$^{-1}$ in the observational profile, and $-10.62\pm 0.15$~km s$^{-1}$ deg$^{-1}$ in the profile from our simulation) is found for this population. Metal-poor, thick disc stars, in turn, show a hot kinematics trend at all latitudes, which results in a flatter gradient with latitude $b$. The consequence of this different response to the bar and b/p bulge results in a metal-rich component that appears kinematically hotter (along the line of sight) than the metal-poor one. Overall, the line-of-sight velocity dispersion-latitude relations observed in the Galactic bulge for metal-poor and metal-rich stars are  both an effect of the intrinsic nature of the Galactic bulge (i.e. mostly secular) and of the angle at which we observe it from the Sun.\\
In this context, we emphasise that extremely metal-rich stars \citep[see also][]{rix24}, defined in this work as stars with [Fe/H] > 0.3,  show the same line-of-sight velocity dispersion-latitude relation as that of the metal-rich stars with less extreme metallicities (i.e. stars with $0 < \rm{[Fe/H]} < 0.3$). In terms of line-of-sight kinematics, these stars are thus not exceptional, but share the same behaviour as the majority of metal-rich stars. This suggests that, as the bulk of the metal-rich population, extremely metal-rich stars are also part of the bar. As for the metal-rich population as a whole, the characteristics of their line-of-sight velocity dispersion-latitude relation (steep increase in $\sigma_{v_{los}}$ with decreasing latitude, and high values of $\sigma_{v_{los}}$ close to the Galactic midplane) suggest that these stars were born in a dynamically cold population and were then trapped in the bar instability.\\ \\

More generally, with this work we confirm that a scenario in which most of the mass in the Galactic bulge (i.e. at [$\textrm{Fe/H]} > -1$) formed out of disc material, where metal-rich and metal-poor populations coexisted and were characterised by different kinematics (the lower the metallicity, the higher the velocity dispersion) is able to reproduce all the trends observed in the Galactic bulge, from the split in the red clump distribution and its dependence on metallicity \citep{ness12, Li_2012, Fragkoudi_2017, debattista17}, to the metallicity distributions and their change with longitude and latitude across the bulge region \citep{fragkoudi_2017b, fragkoudi18}.
Together with the existence of an extended age range for stars in the bulge \citep{bensby13, Ness_2014, Dekany_2015, Haywood_2016}, our results allow for an even more robust anchoring of the origin of the Galactic bulge to that of the stellar populations in the inner disc.

\begin{acknowledgements}
We thank the anonymous referee for useful comments which helped to improve this paper.

This work has made use of the computational resources obtained through the DARI grant A0120410154 (P.I.: P. Di Matteo).

This work has used data from the APOGEE~DR17 catalogue. Funding for the Sloan Digital Sky Survey IV has been provided by the Alfred P. Sloan Foundation, the U.S. Department of Energy Office of Science, and the Participating Institutions. SDSS acknowledges support and resources from the Center for High-Performance Computing at the University of Utah. The SDSS website is \url{www.sdss4.org}. SDSS is managed by the Astrophysical Research Consortium for the Participating Institutions of the SDSS Collaboration including the Brazilian Participation Group, the Carnegie Institution for Science, Carnegie Mellon University, Center for Astrophysics | Harvard \& Smithsonian (CfA), the Chilean Participation Group, the French Participation Group, Instituto de Astrofísica de Canarias, The Johns Hopkins University, Kavli Institute for the Physics and Mathematics of the Universe (IPMU) / University of Tokyo, the Korean Participation Group, Lawrence Berkeley National Laboratory, Leibniz Institut für Astrophysik Potsdam (AIP), Max-Planck-Institut für Astronomie (MPIA Heidelberg), Max-Planck-Institut für Astrophysik (MPA Garching), Max-Planck-Institut für Extraterrestrische Physik (MPE), National Astronomical Observatories of China, New Mexico State University, New York University, University of Notre Dame, Observatório Nacional / MCTI, The Ohio State University, Pennsylvania State University, Shanghai Astronomical Observatory, United Kingdom Participation Group, Universidad Nacional Autónoma de México, University of Arizona, University of Colorado Boulder, University of Oxford, University of Portsmouth, University of Utah, University of Virginia, University of Washington, University of Wisconsin, Vanderbilt University, and Yale University.

The following python libraries were used for this study: \texttt{Astropy} \citep{astropy:2013, astropy:2018, astropy:2022}, \texttt{scipy} \citep{scipy}, \texttt{numpy} \citep{numpy} and \texttt{matplotlib} \citep{matplotlib}.\end{acknowledgements}
\bibliographystyle{aa}
\bibliography{bibliography.bib}

\begin{appendix}
\section{Dispersion profiles for the initial conditions}\label{ic}
\FloatBarrier

\begin{figure}[H]
    \centering
    \includegraphics[width=.5\textwidth]{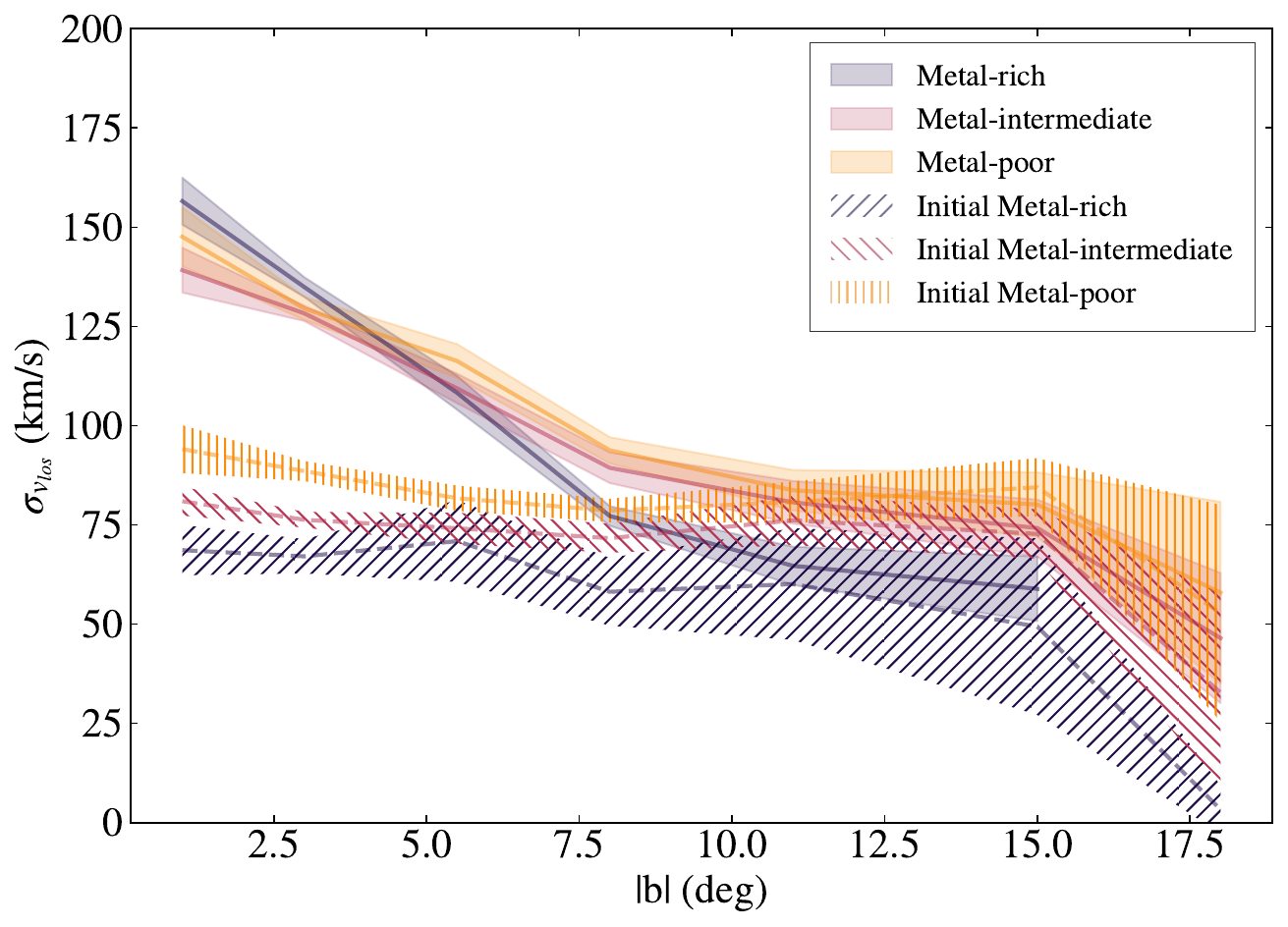}
    \caption{\rev{Line-of-sight velocity dispersion, $\sigma_{v_{los}}$, profiles of the three components in their initial condition (hatched areas and dotted lines) compared to the final profiles (filled areas and solid lines), with the same binning and colour coding used in Fig.~\ref{fig:disp_apogee_sim}. The profiles are plotted for stars with |$\ell$|$<2^\circ$ and |X| $<$ 1.5kpc.}}
    \label{fig:disp_ic}
\end{figure}

\rev{
In Fig.~\ref{fig:disp_ic} we show $\sigma_{v_{los}}$ profiles, similarly to Fig.~\ref{fig:disp_apogee_sim}, but for the metal-rich, metal-intermediate and metal-poor stars after setting the initial conditions of the simulation, that is at time t=0~Gyr. The profiles were computed using bootstrapping, and the hatched areas show the uncertainties associated with the stochastic sampling and the propagated uncertainties caused by the procedure of building a mock catalogue, as described in Appendix~\ref{mock}. At latitudes $|b| \gtrsim 10^\circ$, the metal-rich profile displays high uncertainties, on account of the low star count at heights greater than the scale height of the associated thin disc. The inversion effect is absent from the initial profiles as the velocity dispersion is simply inversely correlated with metallicity. The formation of the b/p bulge generates the dispersion inversion effect seen in the profiles of Fig.~\ref{fig:disp_apogee_sim}, and in the solid lines of Fig.~\ref{fig:disp_ic}, at low latitudes. At higher latitudes, the profiles are left unchanged.
}

\section{Building a mock catalogue}\label{mock}
    \begin{figure}
        \includegraphics[width=.5\textwidth]{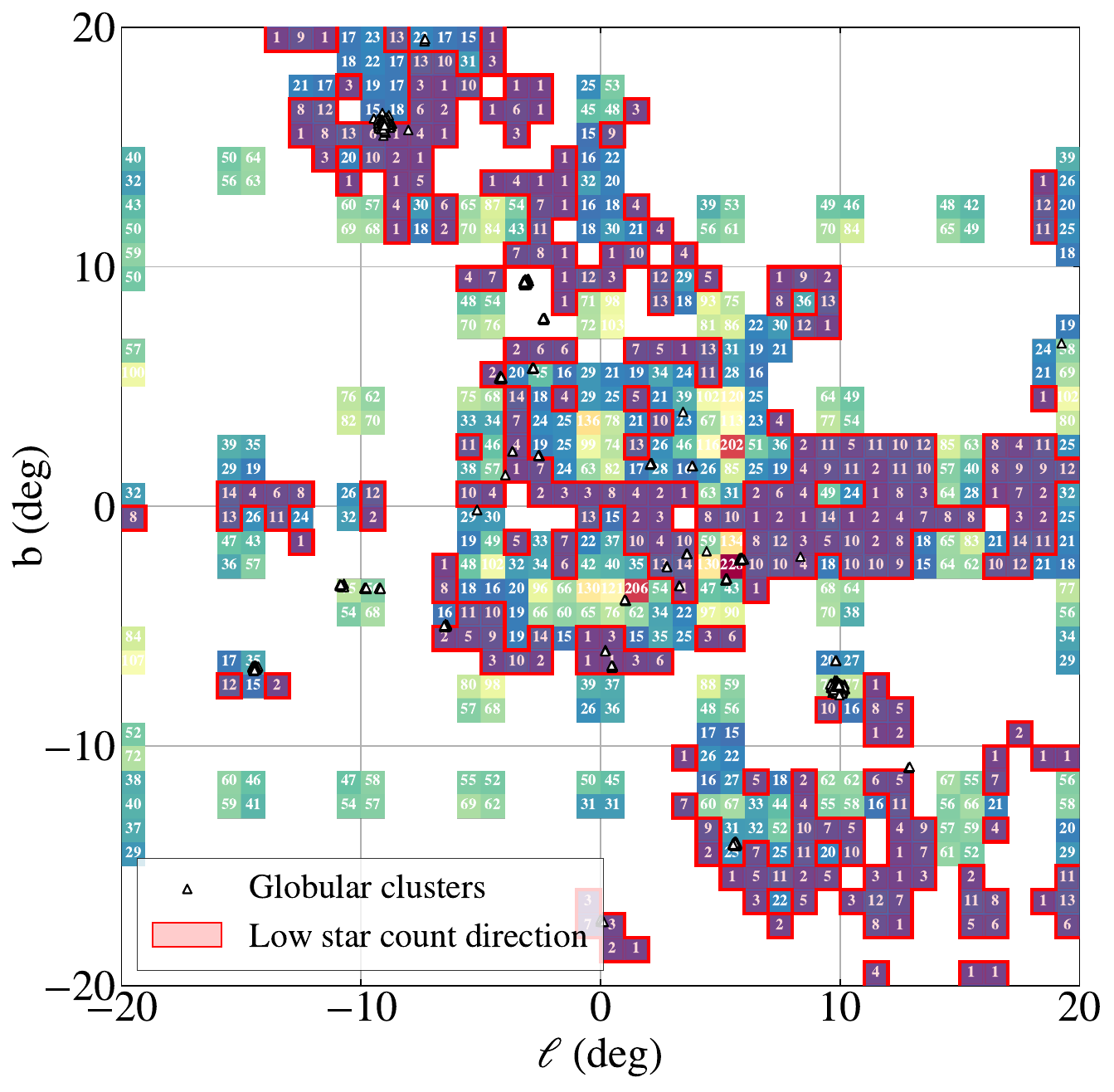}
        \caption{Star count map of our quality sample of APOGEE stars in the Galactic ($\ell$,b) plane. Low-star count regions where no DDF can be constructed are painted in red, and globular clusters stars excluded from our sample are marked with triangle symbols.}
        \label{fig:apogee_lb_distri}
    \end{figure}
    
To compare dispersion profiles built with observational data from APOGEE DR17, we created a mock catalogue from our $N$-body simulation, taking into account observation biases. We first used APOGEE distance uncertainties to scatter the simulation particles along their respective ($\ell$,$b$) direction in order to input a `measure-caused' scatter in our model. We then computed DDFs in each ($\ell$,$b$) direction for the APOGEE bulge stars, and sampled our model particles in such a way that the resulting DDFs reproduce the observational DDFs. In practice, building DDFs is not straightforward as some directions have low star counts; therefore, we used distance uncertainties to smooth the DDFs, making the sampling of our model particles more resilient to low statistics. The star count along each direction is shown in Fig.~\ref{fig:apogee_lb_distri}, where low statistic fields (star counts < 25) are displayed in red. Stars with a globular cluster membership probability higher than 90\% have been excluded and are plotted as triangle symbols. While many fields seem to fall in the low-statistics category, we point out that the star counts sequence along the minor axis near $\ell$=0$^\circ$  -- which is the region of interest for our analysis -- is relatively high (especially at positive latitudes), up to $b$ = 17°. As can be seen from the middle panel of Fig.~\ref{fig:xy_lb}, strong selection biases produce a sparse cover of the ($\ell$,$b$) plane.

\section{Measuring the eccentricity of the bar isodensity contours}\label{eccentricity}
    \begin{figure*}
        \centering
        \begin{minipage}{0.57\textwidth}
            \vspace{-0.2cm}
            \centering
            \includegraphics[width=\hsize]{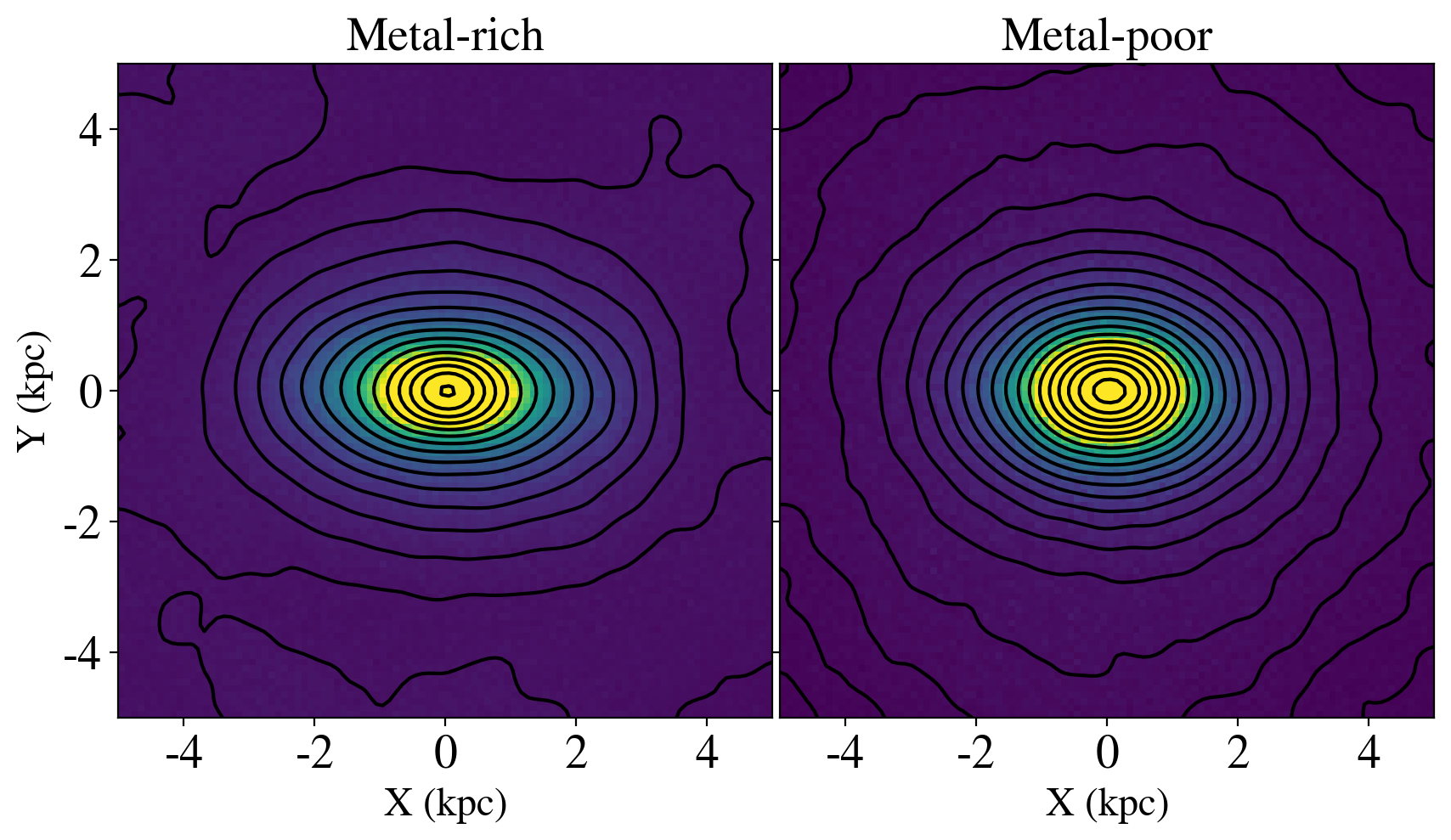}
            \caption{Density maps for the metal-rich (\textit{left panel}) and for the metal-poor (\textit{right panel}) population, represented in a bar-aligned reference frame\rev{ and normalised by the number of particles in each component (D1 and D3)}. Isodensity contours are added as solid lines.}
            \label{fig:orbits}
        \end{minipage}\hfill
        \begin{minipage}{0.415\textwidth}
            
            \centering
            \includegraphics[width=\hsize]{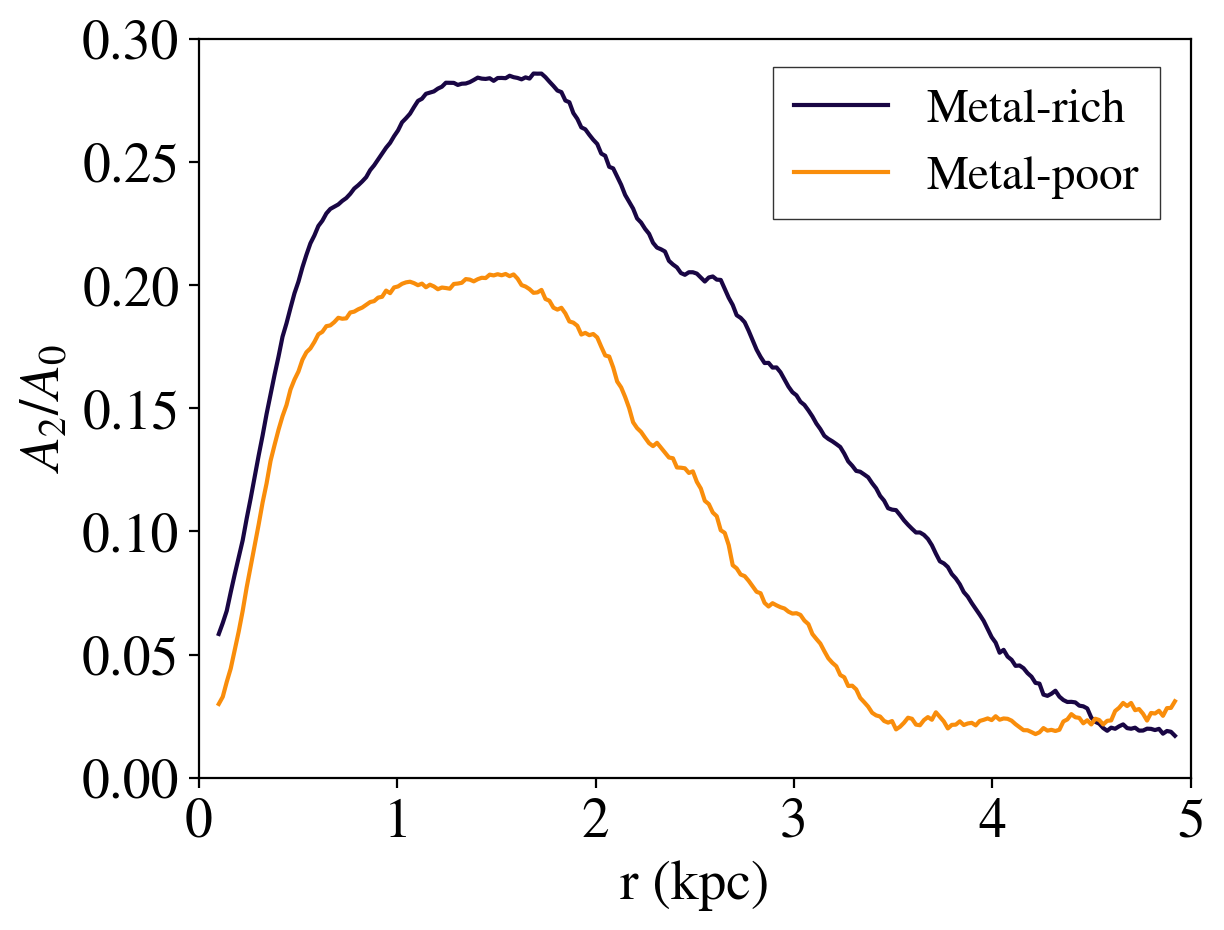}
        \caption{Bar strength of the metal-rich population (purple curve) and metal-poor population (orange curve) against mean radius, as measured by the ratio of the $A_2/A_0$ Fourier coefficients.}
            \label{fig:ecc}
        \end{minipage}
    \end{figure*}
    \label{app:ecc}
    
To quantify the morphological differences between the structures sustained by different populations of stars, we measured the bar strength for the metal-rich and metal-poor components. To this end, we used a standard Fourier coefficient $A_m$ method, computed as follows:
$$a_m = \sum_{i=0}^{N}\cos(m\theta_i)$$
$$b_m = \sum_{i=0}^{N}\sin(m\theta_i)$$
$$A_m = \sqrt{a_m^2 + b_m^2},$$
where the $m=0$ mode is simply the number of particles N, and the $m=2$ mode is a measure of the dipolarity of our sample. With this method, we also have access to the angle of the galactic bar $\Phi$, as $\Phi = atan(b_2/a_2)/2$. We measure a constant 46~km s$^{-1}$ kpc$^{-1}$ bar pattern speed value in the last 1~Gyr. We computed the Fourier coefficients for rings of increasing radii around the GC, and we plot the results against each radius in Fig.~\ref{fig:ecc}, for the metal-poor population in orange and the metal-rich population in purple. We find that the bar strength of metal-rich stars is consistently stronger than their metal-poor counterpart.  Figure~\ref{fig:orbits} shows the density maps of the metal-rich and metal-poor components for the last snapshot of the simulation.  We observe the same overall difference: the metal-rich population having a stronger bar signature than the metal-poor population. 

\section{Dispersion maps in the ($\ell,b$) plane}\label{lb_v}
\FloatBarrier
\begin{figure}[H]
    \includegraphics[width=0.5\textwidth]{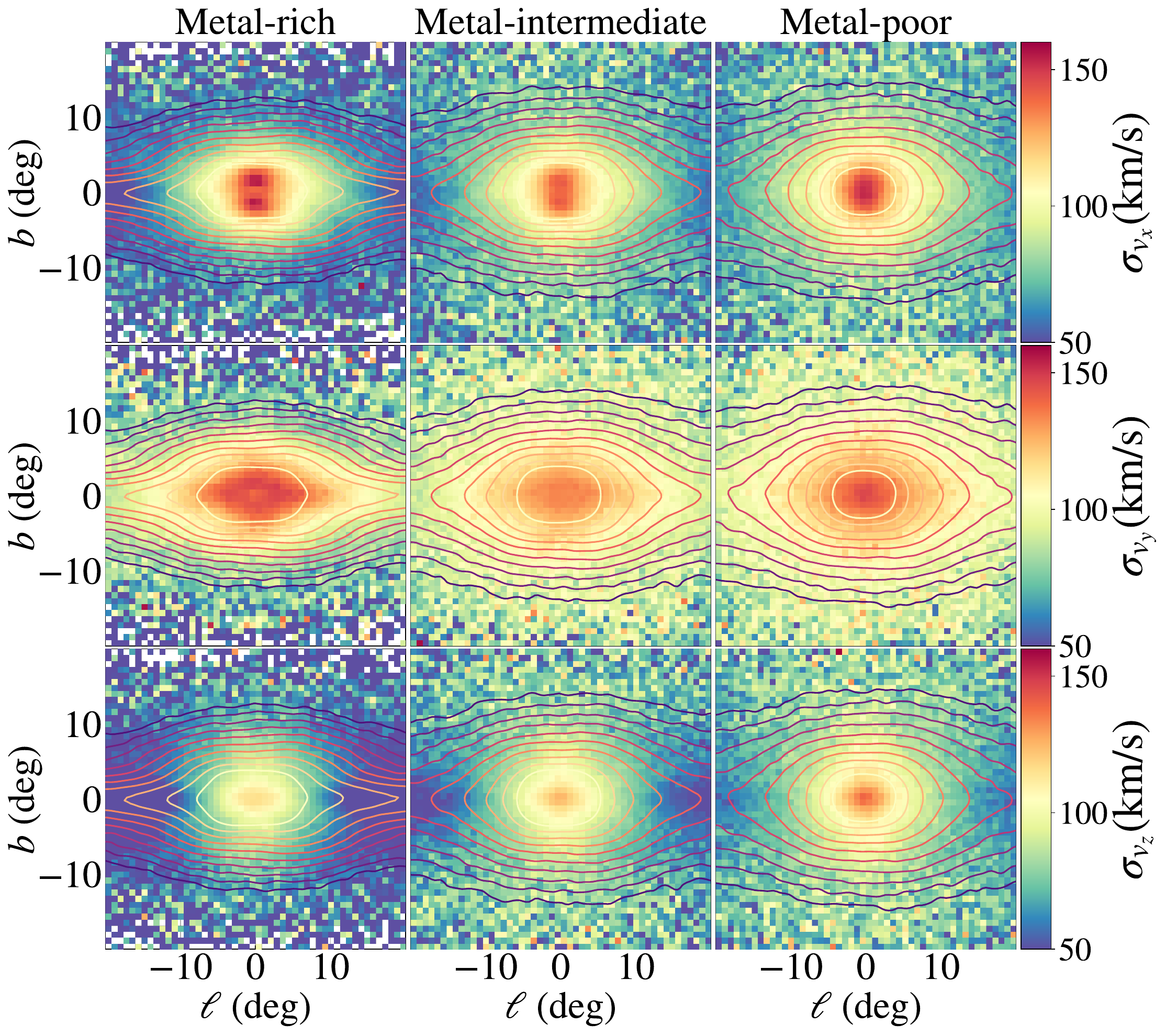}
    \caption{ ($\ell$,$b$) maps of the velocity dispersion $\sigma_v$ maps for the bar-oriented \rev{$\sigma_{v_{Xb}}$} (first row), \rev{$\sigma_{v_{Yb}}$} (second row) and \rev{$\sigma_{v_{Zb}}$} (third row) components. Each population (metal-rich, metal-intermediate and metal-poor) corresponds to a column. Isodensity contours are shown as coloured lines.}
    \label{fig:lb_v}
\end{figure}

\end{appendix}

\end{document}